\documentclass[fleqn,twoside]{article}
\usepackage{espcrc2}

\usepackage{graphicx}
\usepackage[figuresright]{rotating}

\newcommand{\AmS}{{\protect\the\textfont2
  A\kern-.1667em\lower.5ex\hbox{M}\kern-.125emS}}

\title{Belle Computing System}

\author{Ichiro Adachi\address[KEK]{IPNS, KEK, Oho 1-1, Tsukuba, Ibaraki 305-0801, Japan}%
	\thanks{corresponding author.  tel: +81-29-864-5320.  mail address: \tt{ichiro.adachi@kek.jp} },
	Taisuke Hibino\addressmark[KEK],
	Luc Hinz\address[LPHE]{LPHE, EPFL, Lausanne, Dorigny CH-1015, Switzerland},
	Ryosuke Itoh\addressmark[KEK],
	Nobu Katayama\addressmark[KEK],
	Shohei Nishida\addressmark[KEK],
	Fr\'ed\'eric Ronga\addressmark[LPHE]\thanks{present address: IPNS, KEK, Japan}, 
	Toshifumi Tsukamoto\addressmark[KEK] and
	Masahiko Yokoyama\addressmark[KEK] }

\begin{document}

\begin{abstract}
We describe the present status of the computing system in the Belle experiment
at the KEKB $e^+e^-$ asymmetric-energy collider. So far, we have logged more than 160 fb$^{-1}$ of data,
corresponding to the world's largest data sample of 170M $B\bar{B}$ pairs
at the $\Upsilon(4S)$ energy region. A large amount of event data has to be processed
to produce an analysis event sample in a timely fashion. In addition,
Monte Carlo events have to be created to control systematic errors accurately.
This requires stable and efficient usage of computing resources.
Here we review our computing model and 
then describe how we efficiently proceed DST/MC productions in our system.
\vspace{1pc}
\end{abstract}

\maketitle

\section{Introduction}

The Belle experiment\cite{belle-exp} is the $B$-factory project at KEK to study
$CP$ violation in $B$ meson system. The KEKB accelerator\cite{kekb} is 
an asymmetric energy collider
with 8 GeV electron to 3.5 GeV positron, operating at $\Upsilon(4S)$ energies.
The Belle group has been taking data since June 1999 and has logged
more than 160 fb$^{-1}$ of data until December 2003. This means that we
have the largest data sample of $B$ meson pairs around the $\Upsilon(4S)$
energy region in the world.  The KEKB reached its design
luminosity of 10$^{34}$ cm$^{-2}$sec$^{-1}$ on May 2003. 
The KEKB performance is still being improved and we can accumulate integrated luminosity
of about 800 pb $^{-1}$ per day recently.

We have to promptly process all of beam data to provide them for user analyses.
To do this, the DST production should be stable and the computing resources
have to be used in an efficient way. The Monte Carlo( MC ) data should be
generated with statistics large enough to control experimental systematics.
As a result, data size which we should handle is extremely huge
and a mass storage system has to be used to avoid network traffic, 
and data management for entire data sets should be carefully done
without loosing flexibility, for instance,
any modification of data distributions.

Provided a large amount of data sample,
we have published a variety of physics results related to $B$ meson decays, which
is highlighted by the first observation of the large $CP$ violation in
$B$ meson decays\cite{CPVresults}. 
The quick and stable data processing greatly
contributed to this remarkable achievement.

In this paper, we will describe a detail of our computing system
after a brief sketch of Belle software utilities.
In the next section, how we proceed DST as well as MC productions will be mentioned, 
and then summary will be given.

\section{Belle Software}

\subsection{Core Utility}

In the Belle experiment, unique software framework called as
B.A.S.F.( Belle AnalySis Framework ) for all phases in event
processing from online data-taking to offline user analyses has been
employed. This is ``home-made'' core software developed by the Belle
group. In this scheme, each program, written in C$++$, 
is compiled as a shared object,
and it is treated as a module. When one wants to run a program
with this framework, the modules defined in the one's script are
dynamically loaded. 

The data handling is done with the traditional bank system, named as PANTHER,
with a zlib compression capability
In this utility, data transfer between different modules is made and data I/O 
is maniplated. PANTHER is only software to handle our data
in any stage of data processing. 

The typical event data size for rawdata is 35 KB, and 
it is increased up to 60 KB for reconstructed DST data, which
contains all of detector information after unpacking, calibration
and analysis. For user analyses, compact data set (``mini-DST'')
is produced, which is approximately 12 KB for one hadronic event.

\subsection{Reconstruction and Simulation Library}

The reconstruction package is built when major update of programs,
which can affect final physics analysis, is made. Usually it is
built once or twice per year. Once new library is released, we need to
process all of events to produce a consistent data set for analysis.

For MC data,
the detector response is simulated based upon the GEANT3 library\cite{geant}.
Here background events, calculated from beam data, are overlaied onto MC events.
The same reconstruction codes are also applied to MC simulated events.

The detector calibration constants are stored in the database, for which
PostgreSQL\cite{postgres} is adopted in the Belle experiment. Two database servers
are running in our computing system, where one is for public usage and
the other for DST/MC productions.  The contents of each
server are periodically mirrored to the other, and
a backup tar file for the contents of the original database server 
is archived once per week.
For linux users, one can start up own database server in her/his PC, according to
a Belle instruction. User, if necessary, can download original database contents from  
a KEK B computer for private purpose.

\section{Belle Computing System}

\begin{figure}[t]
\centering
\includegraphics[width=65mm]{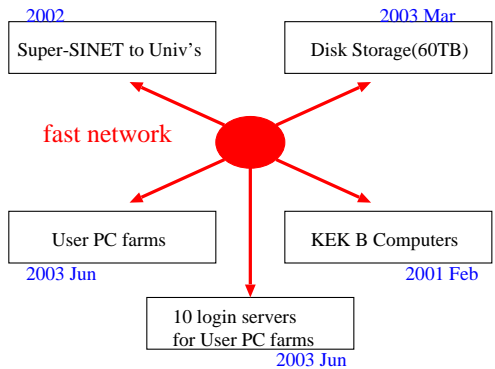}
\caption{Computing model overview.} 
\label{fig:comp_model}
\end{figure}

\begin{figure*}[t]
\centering
\includegraphics[width=135mm]{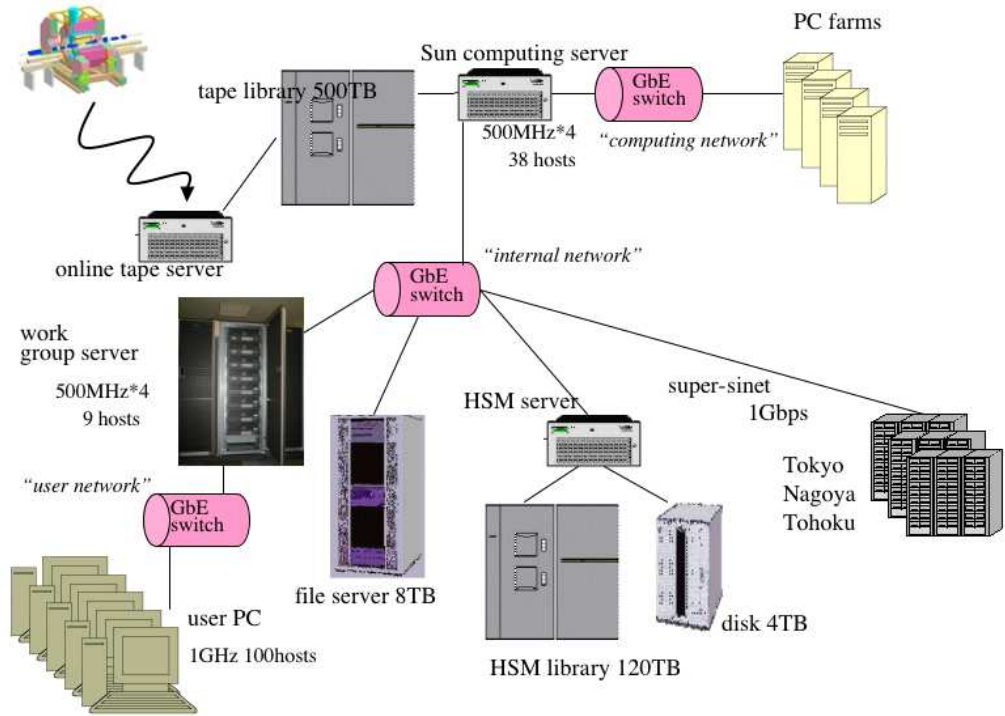}
\caption{Belle computing system.} 
\label{fig:belle_computing}
\end{figure*}

\subsection{Computing Model Overview}

Figure~\ref{fig:comp_model} indicates an overview of our computing model for the Belle experiment. 
As can be seen, it comprises of the four major elements. 
The first one is the KEK B Computers
which has been operated since February 2001\cite{kekcomp}. This has been a 
principal system, where data processing as well as analyses have been performed.
In addition to this, we have equipped a 60 TB disk storage area March 2003.
In this space, more beam and MC data can be kept.
As for a network inside Japan, Super-SINET link has been available since 2002\cite{supersinet}.
To enrich CPU's for user analyses, PC farms dedicated to analysis jobs
have been installed in June 2003. These components are interlinked each other
with a fast network of 1$\sim$10 Gbps. 

\begin{table*}
\caption{\small A summary of CPU's for the KEK B computers. }
\label{tb:KEKB-CPU}
\begin{tabular}{cccccc} \hline
host              & processor &  clock    & \#CPU's & \#nodes    \\ \hline  
computing server  & sparc     & 500MHz    &    4    &      38    \\ 
work group serve  & sparc     & 500MHz    &    4    &       9    \\ 
user PC           & P3        & 1GHz      &    1    &     100    \\ \hline
\end{tabular}
\end{table*}

\subsection{KEK B Computers}

The KEK B computers are schematically shown in Figure~\ref{fig:belle_computing}. 
This system consists of the three cardinal links. 
The first one is called ``computing network'', which
interconnects the 38 Sun servers with the PC farms, which will be described below. 
These Sun hosts are operated as a batch job system controlled by the LSF scheduler.
The computing network is also attached to the DTF2 tape robotic device with 500 TB total
volume which is used for rawdata and DST data storage.
The next network links the 9 work group servers of the Sun hosts to the storage
devices, which consists of the 8 TB file server and the hierarchy mass storage( HSM )
of 120 TB capacity with the 4 TB staging disk. The work group servers are connected
via the ``user network'' to 100 PC's for users. With this network, user can login from his/her
PC to one of the work group servers, from which one can submit a batch job to the 38 Sun compute
servers. Table~\ref{tb:KEKB-CPU} summarizes the CPU's allocated in the KEK B computers.

\subsection{PC farm}

More and more data have been accumulated, more and more computing needs
for event processing have been seriously arising. To fulfill this requirements,
a bunch of PC's has been installed and connected into this existing network
by considering the best usage without making a bottleneck. Table~\ref{tb:PC-CPU}
tabulates the PC farm CPU's in our system. As one can see, our PC farms are heterogeneous 
from various vendors. The best choice at the moment to get new PC's 
usually made us to purchase from a variety of vendors, and it effectively reduces cost for CPU's.

In all PC's, linux utilities as well as the Belle library packages have been loaded and
we can use them as clusters of seamless PC systems to process event data. 

\begin{table*}
\caption{\small A breakdown of the PC farm CPU's. }
\label{tb:PC-CPU}
\begin{tabular}{cccccc} \hline
vendor & processor &  clock    & \#CPU's & \#nodes  & total CPU    \\ \hline  
Dell   & P3        & 500MHz    &    4    &      16  &     32GHz    \\ 
Dell   & P3        & 550MHz    &    4    &      20  &     44GHz    \\ 
Compaq & P3        & 800MHz    &    2    &      20  &     32GHz    \\ 
Compaq & P3        & 933MHz    &    2    &      20  &     37GHz    \\ 
Compaq & Intel Xeon& 700MHz    &    4    &      60  &    168GHz    \\ 
Fujitsu& P3        & 1.26GHz   &    2    &     127  &    320GHz    \\ 
Compaq & P3        & 700MHz    &    1    &      40  &     28GHz    \\ 
Appro  & Athlon    & 1.67GHz   &    2    &     113  &    377GHz    \\ 
NEC    & Intel Xeon& 2.8GHz    &    2    &      84  &    470GHz    \\ \hline
Total  &           &           &         &     500  &   1508GHz    \\ \hline 
\end{tabular}
\end{table*}

The primary purpose to add PC farms is that we have to increase CPU power for
DST and MC productions. In 2003, we expanded usage for our PC's by
releasing new PC farms for user analyses, as a possible solution for
ever increasing users' demand. The 10 login servers have been 
arranged with 6 TB local disk area, where
user histogram files and analysis codes are located. From each login server,
job can be submitted to the user PC farm consisting of 84 PC's with
dual Xeon 2.8 GHz CPU's. All PC's are managed by the LSF queuing utility.
From user PC farms, beam and MC data samples, which are stored in 
the 60 TB disk mentioned previously, can be accessed. 

\subsection{Super-SINET at Belle}

A fast 1 Gbps network dedicated to academic activities, Super-SINET\cite{supersinet}, 
has been available between KEK and major Japanese universities. This link enables us to copy
a bulk of beam and MC data from KEK to other institutions and vice versa. 
Moreover, computing resources can be shared as seamless system using Super-SINET. 
For example, one disk connected to PC at Nagoya university can be mounted to the KEK computer
via NFS as if it were located inside the KEK site. Then, output data can be written
onto the disk directly from the KEK computer. It is possible to make efficient and full
use of total resources collaboration-wide.

\subsection{Data Management}

A large volume of data is comprised of more than 30 K data files including beam and MC events.
Basically each user has to go through all of these files to obtain final physics
results. So, it is very important to notify all of file locations to users for their analyses.
These data files are distributed
over a bunch of storage disks and sometimes data files have to be moved for various
reasons such as disk failure and so on. From administrative point of view, 
it is necessary to maintain flexibility in data management despite of any change of the file locations.
To solve this, we have registered all of the attributes of data files such as locations, data type 
and number of events into our database, and they serves as ``metadata'' to access
actual beam data. The central database contents for data file information is maintained 
and updated whenever new data is available. 
The web-based interface between database and users is prepared
to extract necessary information. In user's batch job, inquiry to the database
is automatically issued and user can easily analyze event data without knowing actual file
locations. 

\section{Data Processing}

\begin{figure*}[t]
\centering
\includegraphics[width=135mm]{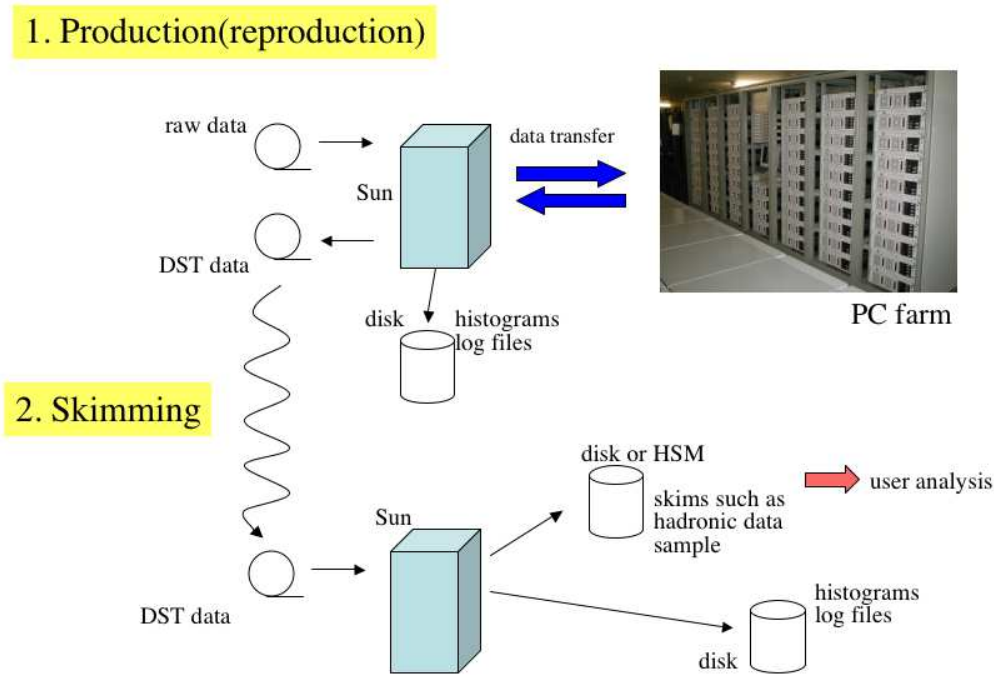}
\caption{A schematic drawing of the DST production scheme.} 
\label{fig:prod_scheme}
\end{figure*}

\subsection{Beam data production}

\begin{figure}[t]
\centering
\includegraphics[width=65mm]{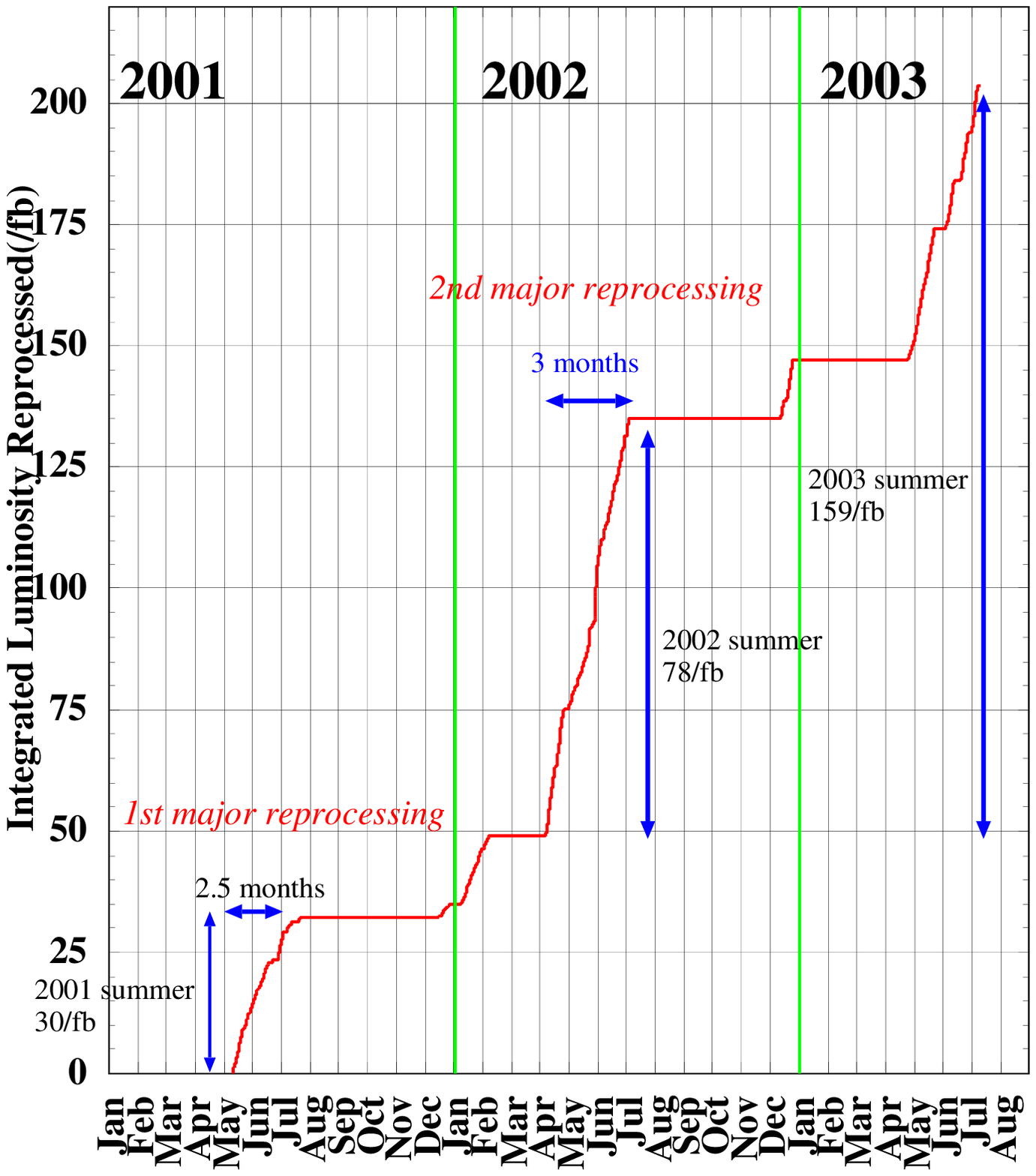}
\caption{History of reprocessing from 2001.} 
\label{fig:reprocess}
\end{figure}

The scheme of the DST production is shown in Figure~\ref{fig:prod_scheme}. In the first step
of the DST production, one of Sun
compute host is assigned as a tape server, and two DTF2 tapes, one is for
raw data and the other for DST data, are mounted. Then, raw data are read from the tape
and are distributed over PC nodes. In each PC node, event processing is performed
in the B.A.S.F. framework. After the reconstruction is made, event data are sent back to
the tape server, where data are written onto the tape as DST. 

The next step, the event skimming, is carried out in such a way that DST data are again read 
from the DST2 tape at the Sun compute server, and an appropriate selection 
criteria is imposed onto the data. In case that an event satisfies selection
conditions, it is saved onto disk as a skimmed event. Each event is examined
by a set of selections such as $\mu$-pair event for a detector study and
hadronic event for a physics analysis.

In our system, we can process about 1 fb$^{-1}$ of beam data per day with 40 PC nodes of
quad Intel Xeon 700 MHz CPU's each, and it can be increased up to 2 fb$^{-1}$ per day
by adding PC nodes. Figure~\ref{fig:reprocess} shows a history of our beam data processing from 2001.
In 2001, we completed the first entire reprocessing with a single version of
the reconstruction package, providing 30 fb$^{-1}$ of data sample for 2000 summer 
conferences. In 2002, major update of the software was made in April 2002 and
the second reprocessing for 78 fb$^{-1}$ of data using this library was
performed from April to June. Last year, we added another 80 fb$^{-1}$
by reprocessing them, amounting to 159 fb$^{-1}$ of total beam data.

\subsection{MC production}

\begin{figure}[t]
\centering
\includegraphics[width=65mm]{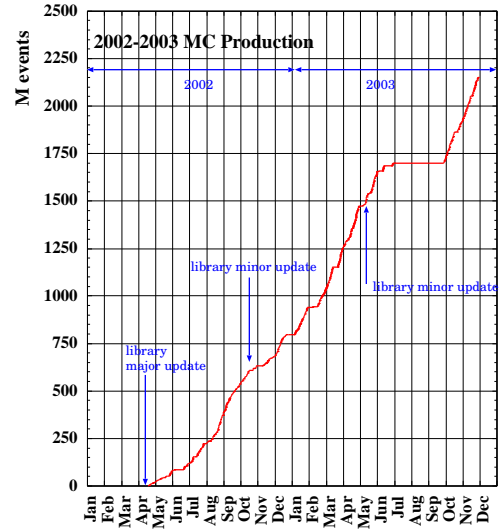}
\caption{MC production done in 2002 and 2003.} 
\label{fig:MCprocess}
\end{figure}

We have three types of MC samples, $B^0\bar{B^0}$, $B^+B^-$ and continuum events.
They are produced on a run-by-run basis, where each MC sample file corresponds to
each beam data file. After data production for one run beam file is done, run
dependent information like beam interaction profile and background hit rate
are calculated immediately and, based upon these information, MC sample data
corresponding to the beam run file is created for three types.  
Beam background hits are overlaied with MC generated data to mimic actual events
as precisely as possible. Figure~\ref{fig:MCprocess} shows how we have produced MC samples
recent two years. We have produced 2.2 billion events in total by
the end of 2003, which is equivalent to 3 times larger statistics
for 159 fb$^{-1}$ real beam data.

Another aspect of the MC production is contribution from the remote institutes 
outside KEK. Approximately one quater of MC events have been produced at remote sites
and they are transferred to KEK via network.

\section{Summary and Plan}

The Belle computing system has been successfully working and 
we have processed a bulk of beam data of more than 250 fb$^{-1}$ 
so far. For MC data, 2.2 billion events have been produced, which corresponds
to more than 3 times larger statistics than real data.
Those data have been managed via our database and this scheme
allows us to provide the data sets steadily and flexibly.

In 2003 summer, new silion vertex detector has been installed 
it is expected that this detector expands our tracking and vertexing
capability further. At present, reconstruction algorithm based upon
new Belle configuration is being developed. After it is
completed, beam data in 2003 autumn runs will be processed. To do this,
we are planninng to double our processing power for beam data by adding
more CPU's and storage devices.

\end{document}